\documentclass[twocolumn,showpacs,preprintnumbers,amsmath,amssymb,aps,prl]{revtex4-1}

\usepackage{graphicx}
\usepackage{dcolumn}
\usepackage{bm}
\usepackage{color}

\usepackage[hypertex,dvipdfmx]{hyperref}
\hypersetup{colorlinks=true,breaklinks,linkcolor=blue,urlcolor=blue,citecolor=blue}

\begin{document}

\preprint{}

\title{Competing $d$-Wave and $p$-Wave Spin-Singlet Superconductivities\\ in the Two-Dimensional Kondo Lattice}

\author{Junya Otsuki}
\affiliation{%
Department of Physics, Tohoku University, Sendai 980-8578, Japan
}%

\date{\today}

\begin{abstract}
The Kondo lattice model describes a quantum phase transition between the antiferromagnetic state and heavy-fermion states.
Applying the dual-fermion approach, we explore possible superconductivities emerging due to the critical antiferromagnetic fluctuations.
The $d$-wave pairing is found to be the leading instability only in the weak-coupling regime.
As the coupling is increased, we observe a change of the pairing symmetry into a $p$-wave spin-singlet pairing.
The competing superconductivities are ascribed to 
crossover between small and large Fermi surfaces,
which occurs with formation of heavy quasiparticles.
 
\end{abstract}


\maketitle

Heavy-fermion superconductivities in rare-earth and actinide materials commonly appear near antiferromagnetic (AFM) or ferromagnetic quantum critical point (QCP).
This situation resembles high-$T_{\rm c}$ cuprates, where $d$-wave superconductivity ($d$-SC) with the gap symmetry $\phi_{\bm{k}} \propto \cos k_x - \cos k_y$ is realized due to AFM fluctuations of $\bm{q}=(\pi,\pi)\equiv\bm{Q}$.
This similarity suggests a possibility of the $d$-SC also in heavy-fermion systems near AFM QCP. 
However, since the microscopic origin of the magnetism is different between $d$- and $f$-electron systems, pairing interactions derived from the magnetic fluctuations may be dissimilar, resulting in a different type of superconductivity in $f$-electron systems.

Motivated by the above idea, we consider the following simplified situation to focus on a difference in the microscopic interactions. 
We first suppose a tight-binding band on the square lattice, $\epsilon_{\bm{k}} = -2(\cos k_x + \cos k_y)$, which exhibits a perfect nesting at $\bm{q}=\bm{Q}$ at half-filling.
In the case of the cuprates, the Hubbard interaction $U$ gives rise to the AFM long-range order, and carrier doping leads to the $d$-SC.
In heavy-fermion systems, on the other hand, the nesting of the conduction band leads to the AFM ordering of localized $f$-electrons through the RKKY interaction.
Our question now is whether the $d$-SC is realized in this case as well by tuning the system toward QCP, and if not, which type of pairing is favored?

For this purpose, we explore possible unconventional superconductivities in the two-dimensional Kondo lattice model. 
The Hamiltonian reads
\begin{align}
{\cal H}
= \sum_{\bm{k}\sigma} \epsilon_{\bm{k}} c_{\bm{k}\sigma}^{\dag} c_{\bm{k}\sigma}
+ J \sum_i \bm{S}_i \cdot \bm{s}_i,
\label{eq:Hamil}
\end{align}
where $\bm{s}_i=(1/2) \sum_{\sigma\sigma'} c_{i\sigma}^{\dag} \bm{\sigma}_{\sigma\sigma'} c_{i\sigma'}$.
The number of lattice sites, $N$, is taken to be
$N=L^2$ with $L=32, 64, 128$.
Various numerical methods have been applied to the Kondo lattice with central attention on the magnetic properties~\cite{Assaad99,Watanabe07,Martin08,Misawa13,Li-arXiv,Peters-arXiv}. 
Further elaborate calculations are necessary to elucidate pairing fluctuations lying near the magnetic instability~\cite{Asadzadeh13}.

In order to address superconductivity in the Kondo lattice, we need an approximate method which, at least, captures Kondo physics and AFM fluctuations around QCP.
The dynamical mean-field theory (DMFT) takes full account of local correlations~\cite{Georges96}, and hence of Kondo physics.
Indeed, formation of heavy quasiparticles was traced at finite temperatures~\cite{Otsuki09a},
and a quantum phase transition between the Kondo paramagnet and the AFM state was derived~\cite{Peters07, Otsuki09c}.
We work on an extension of the DMFT to describe influence of critical AFM fluctuations on heavy quasiparticles.

The dual-fermion approach provides a way to perform diagrammatic expansion around the DMFT~\cite{Rubtsov08,Rubtsov09}.
In particular, inclusion of ladder-type diagrams as in the fluctuation exchange (FLEX) approximation describe long-range correlations, which lead to paramagnon excitations~\cite{Hafermann09,Hafermann-book}, the $d$-SC in the Hubbard model~\cite{Otsuki14a}, and the correct critical exponents in the Falikov-Kimball model~\cite{Antipov14}.
Further advantage of the dual-fermion method as compared with related theories~\cite{Kusunose06,Toschi07,Slezak09,Taranto14} is its applicability: The framework is easily applied to the Kondo lattice model as presented below.

{\em Formalism.}
In the path integral representation using Grassmann numbers, the partition function of the model~(\ref{eq:Hamil}) is written as
$Z=\int \prod_i [{\cal D} (c_i^{*} c_i) {\cal D}\bm{S}_i] e^{-{\cal S}}$
with the action
\begin{align}
{\cal S}
= \sum_i {\cal S}^{\rm imp}_{i} 
+ \sum_{\omega\bm{k}\sigma} (\epsilon_{\bm{k}} -\Delta_{\omega}) c_{\omega\bm{k}\sigma}^{*} c_{\omega\bm{k}\sigma}.
\end{align}
Here ${\cal S}^{\rm imp}$ is the local part defined by
\begin{align}
{\cal S}^{\rm imp}_{i}
=-\sum_{\omega\sigma} (i\omega+\mu-\Delta_{\omega}) c_{\omega i\sigma}^{*} c_{\omega i\sigma}
+ J \bm{S}_{i} \cdot \bm{s}_{i}.
\end{align}
The dual-fermion transformation is performed on the kinetic-energy term described by the second term of ${\cal S}$.
On the other hand, the difference between the Kondo lattice model and the Hubbard model lies in the local part ${\cal S}^{\rm imp}$.
Therefore, the procedure deriving the dual-lattice action in the Hubbard model can be directly applied to the Kondo lattice, 
leaving the difference in the local interactions.
Introducing dual fermions $d$ and formally integrating out the variables $c$ and $\bm{S}$ at each site, we arrive at the lattice model
$\tilde{Z}=\int \prod_i {\cal D}(d_i^* d_i) e^{-\tilde{\cal S}}$ 
which is written only with the dual fermions~\cite{Rubtsov08,Rubtsov09}
\begin{align}
\tilde{\cal S}
&=\sum_{\omega\bm{k}\sigma} (-\tilde{G}_{\omega\bm{k}}^0)^{-1} d_{\omega\bm{k}\sigma}^{*} d_{\omega\bm{k}\sigma}
\nonumber \\
&- \frac{1}{4} \sum_{1234} \gamma_{\omega_1 \omega_2 \omega_3 \omega_4}^{\sigma_1 \sigma_2 \sigma_3 \sigma_4}
d_{\omega_1 \bm{k}_1 \sigma_1}^{*} d_{\omega_2 \bm{k}_2 \sigma_2}^{*} d_{\omega_3 \bm{k}_3 \sigma_3} d_{\omega_4 \bm{k}_4 \sigma_4}.
\label{eq:S-dual}
\end{align}
Information of the local interaction enters in $\tilde{G}^0$ and $\gamma$.
The propagator $\tilde{G}^0$ is defined by
$\tilde{G}_{\omega\bm{k}}^0=(g_{\omega}^{-1}+\Delta_{\omega}-\epsilon_{\bm{k}})^{-1} - g_{\omega}$
with 
$g_{\omega}=-\langle c_{\omega i \sigma} c_{\omega i \sigma}^* \rangle_{\rm imp}$ being the local Green's function in the system described by ${\cal S}^{\rm imp}_{i}$.
The interaction coefficient $\gamma$ corresponds to the vertex part in the impurity system:
$\gamma_{1234}
= [ \langle c_1 c_2 c_3^* c_4^* \rangle_{\rm imp}
- g_1 g_2 ( \delta_{14} \delta_{23} - \delta_{13} \delta_{24}) ]
/ (T g_{1} g_{2}  g_{3} g_{4})$
where we used abbreviations such as $1 \equiv(\omega_1, \sigma_1)$.
In deriving Eq.~(\ref{eq:S-dual}), we neglected many-body interactions which involve more than three particles. Validity of this approximation was confirmed numerically~\cite{Hafermann09} and also supported in terms of $1/d$ expansion~\cite{Otsuki14a}.
We compute $g_{\omega}$ and $\gamma$ using the continuous-time quantum Monte Carlo (QMC) method~\cite{Gull11} applied to the Kondo impurity model~\cite{Otsuki07}.

\begin{figure}[tb]
	\begin{center}
	\includegraphics[width=\linewidth]{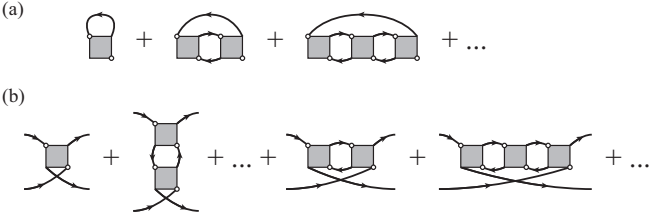}
	\end{center}
	\caption{Diagrams for (a) the dual self-energy $\tilde{\Sigma}$ and (b) the pairing vertex $\Gamma^{\rm PP}$. The lines and squares represent $\tilde{G}$ and $\gamma$, respectively.}
	\label{fig:diagrams}
\end{figure}

Our task is now to evaluate the dual self-energy 
$\tilde{\Sigma}_{\omega\bm{k}}$, 
which is connected to the original Green's function $G_{\omega\bm{k}}$ by the exact formula~\cite{Rubtsov08,Rubtsov09},
$G_{\omega\bm{k}}=[(g_{\omega} + g_{\omega} \tilde{\Sigma}_{\omega\bm{k}} g_{\omega})^{-1} + \Delta_{\omega} - \epsilon_{\bm{k}}]^{-1}$.
We take FLEX-type diagrams in the particle-hole channel expressed in Fig.~\ref{fig:diagrams}(a) to take account of critical AFM fluctuations~\cite{Hafermann09,Hafermann-book,Otsuki14a}~\footnote{The ladder approximation is sufficient for our purpose of revealing pairing symmetries induced by the critical AFM fluctuations. More complicated diagrams need to be included to address, e.g., feedback from pairing fluctuations, which is beyond the scope of this paper.}.
After $\tilde{\Sigma}_{\omega\bm{k}}$ is obtained,
we update the hybridization function $\Delta_{\omega}$ and solve again the effective impurity problem until the self-consistency condition
$\sum_{\bm{k}}\tilde{G}_{\omega\bm{k}}=0$
is fulfilled.

{\em QCP.}
We first identify the quantum phase transition between Kondo insulator and AFM insulator at half-filling. 
According to a lattice QMC calculation~\cite{Assaad99}, the critical coupling is estimated at $J_{\rm c}^{\rm QMC}=1.45 \pm 0.05$.
Comparing with this value, we shall check accuracy of our approximation.

\begin{figure}[tb]
	\begin{center}
	\includegraphics[width=0.9\linewidth]{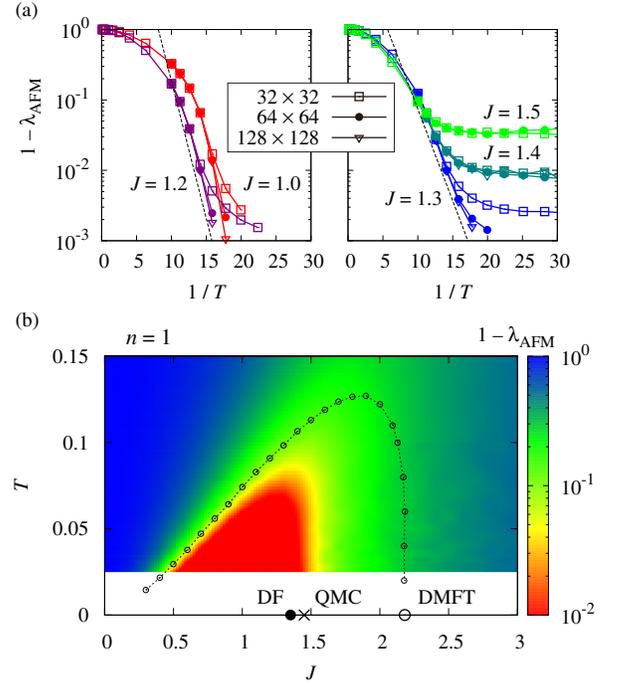}
	\end{center}
	\caption{(Color online) (a) The leading eigenvalue $\lambda_{\rm AFM}$ for the AFM fluctuations at half-filling, $n=1$. The dashed lines indicate the critical behavior $1-\lambda_{\rm AFM} \propto e^{-\beta \Delta}$. (b) An intensity plot of $\lambda_{\rm AFM}$ in the $J$-$T$ plane. The points indicate QCP estimated by QMC, DMFT, and our approximation (DF). The dashed line shows the phase boundary computed in DMFT.}
	\label{fig:AFM}
\end{figure}

In the dual-fermion approach, the AFM fluctuations may be observed through eigenvalues of the matrix
$A_{\omega\omega'}=-(1/N)\sum_{\bm{k}} \tilde{G}_{\omega\bm{k}} \tilde{G}_{\omega,\bm{k}+\bm{Q}} \gamma_{\omega\omega'}^{\rm sp}$~\cite{Otsuki14a}.
Here, $\gamma_{\omega\omega'}^{\rm sp}$ denotes the spin channel of the local vertex, 
$\gamma_{\omega\omega'}^{\rm sp} \equiv \gamma^{\uparrow \uparrow \uparrow \uparrow}_{\omega \omega'\omega'\omega}-\gamma^{\uparrow \downarrow \downarrow \uparrow}_{\omega \omega'\omega'\omega}$,
The transition takes place when the largest eigenvalue $\lambda_{\rm AFM}$ exceeds 1.
Because of the Mermin-Wagner theorem~\cite{Mermin-Wagner}, the AFM transition is forbidden at $T \neq 0$ and $\lambda_{\rm AFM}$ follows the critical behavior $1-\lambda_{\rm AFM} \propto e^{-\beta \Delta}$ when the ground state is AFM~\cite{Chakravarty88,Hasenfratz91}. On the other hand, if the AFM is suppressed by the Kondo effect, $\lambda_{\rm AFM}$ becomes constant at low temperatures. To distinguish these behaviors, we plot $1-\lambda_{\rm AFM}$ as a function of $1/T$ in Fig.~\ref{fig:AFM}(a).
These plots demonstrate that the ground state is AFM for $J\leq1.2$ and paramagnetic for $J\geq1.4$.
The data at $J=1.3$ show a clear size dependence, which is common to the AFM ground state.
Therefore, we conclude that $J=1.3$ is in the AFM region, and the quantum phase transition is estimated at $J_{\rm c}^{\rm DF}=1.35 \pm 0.05$.
This value agrees with the QMC result within 10\%, and shows considerable improvement from the DMFT result $J_{\rm c}^{\rm DMFT} \fallingdotseq 2.18$.
These estimations of $J_{\rm c}$ are summarized in Fig.~\ref{fig:AFM}(b) together with an intensity plot of the AFM fluctuations.
The critical region, e.g., $1-\lambda_{\rm AFM} \lesssim 10^{-2}$, forms a dome which is similar to the phase boundary in the DMFT.

{\em Superconductivity.}
We now compute superconducting instabilities near half-filling.
We solve the linearized Bethe-Salpeter equation for pairing correlations,
$-(T/N)\sum_{k'}\tilde{G}_{k} \tilde{G}_{-k}\Gamma_{kk'}^{\rm PP} \phi_{k'} = \lambda_{\rm SC} \phi_{k}$,
where $k=(\omega,\bm{k})$.
For the irreducible vertex part $\Gamma_{kk'}^{\rm PP}$, we take account of ladder-type vertex corrections in Fig.~\ref{fig:diagrams}(b)
to describe pairing interactions mediated by AFM fluctuations~\cite{Otsuki14a}. 
As in the case of AFM ordering, the superconducting transition can be detected by the condition $\lambda_{\rm SC}=1$.
We computed the leading eigenvalues $\lambda_{\rm SC}$ for all types of pairing symmetry, i.e., 2 spin channels (singlet/triplet) and 5 spatial symmetries classified according to the irreducible representations of the point group D$_{4}$.

\begin{figure}[tb]
	\begin{center}
	\includegraphics[width=0.9\linewidth]{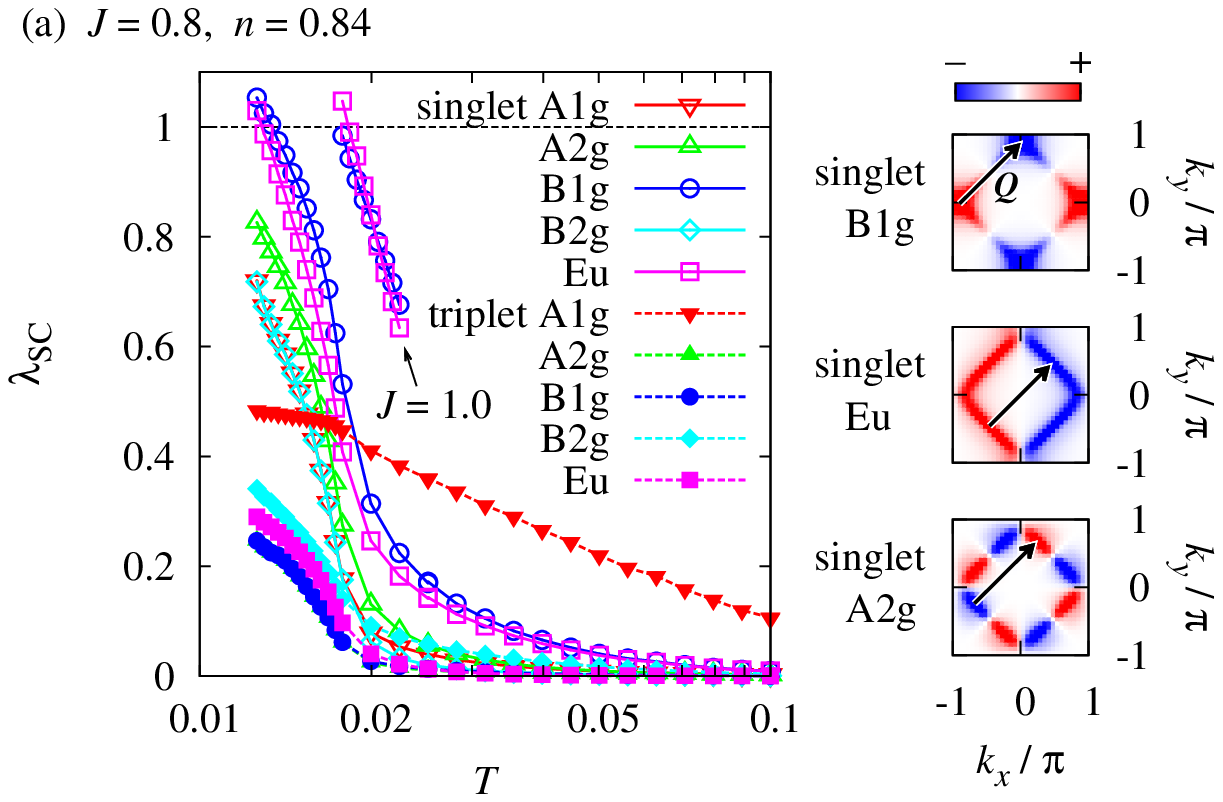}
	\includegraphics[width=0.9\linewidth]{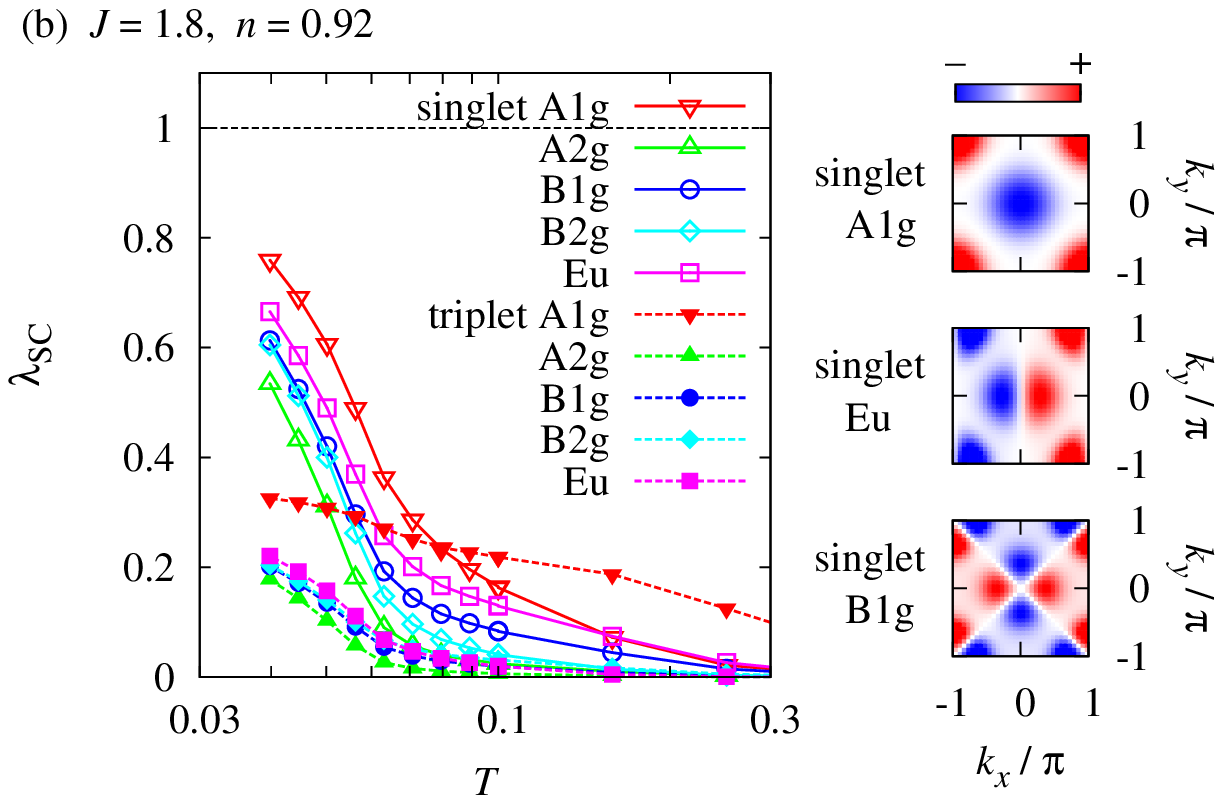}
	\end{center}
	\caption{(Color online) Temperature dependence of the leading eigenvalues $\lambda_{\rm SC}$ for pairing fluctuations. The intensity plots show the corresponding eigenfunctions $\phi_{\omega_0 \bm{k}}$ in the momentum space for the three largest fluctuations. The arrows indicate the nesting vector $\bm{Q}\equiv(\pi,\pi)$.
	(a)~$J=0.8$, $n=0.84$ and (b)~$J=1.8$, $n=0.92$. 
	The system size is $N=32^2$.}
	\label{fig:SC-eigen}
\end{figure}

\begin{figure}[tb]
	\begin{center}
	\includegraphics[width=0.9\linewidth]{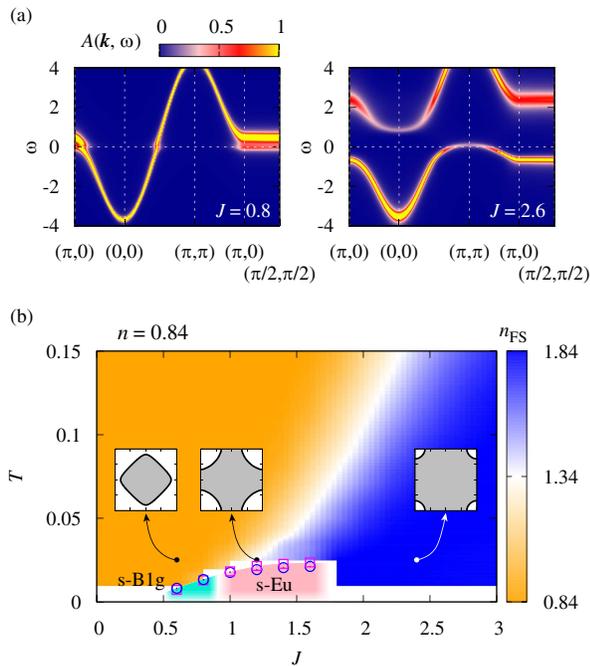}
	\end{center}
	\caption{(Color online) (a) The single-particle excitation spectrum $A(\bm{k},\omega)$ for $n=0.84$ and $T=0.056$. (b) Superconducting transition temperatures of singlet-B$_{\rm 1g}$ (circle) and singlet-$E_{\rm u}$ (square) symmetries. The intensity plot shows the Fermi-surface area $n_{\rm FS}$ defined in Eq.~(\ref{eq:n_FS}). The Fermi-surface structure is plotted for three representative parameters.}
	\label{fig:FS-area}
\end{figure}

Figure~\ref{fig:SC-eigen}(a) shows temperature dependence of $\lambda_{\rm SC}$ for $n=0.84$ in the weak-coupling regime, $J=0.8<J_{\rm c}$.
The leading instability is the spin-singlet pairing with B$_{\rm 1g}$ symmetry as expected. 
However, other symmetries in the spin-singlet channel show comparable enhancement in contrast to the Hubbard model~\cite{Otsuki14a}.
In particular, the E$_{\rm u}$ pairing is nearly degenerate with $B_{\rm 1g}$ and becomes the leading instability in $J=1.0$ as also plotted in Fig.~\ref{fig:SC-eigen}(a).
We note that the E$_{\rm u}$ pairing is an odd function in the frequency domain as well as in the momentum space~\cite{Balatsky92}.
The momentum dependence of the eigenfunctions $\phi_{\omega_0 \bm{k}}$ at the lowest Matsubara frequency $\omega_0=\pi T$ is plotted in Fig.~\ref{fig:SC-eigen}(a) for the three leading fluctuations. 
A common feature is found: The nesting vector $\bm{Q}\equiv(\pi,\pi)$ connects strong-intensity regions with opposite signs.
It indicates that those pairings are due to AFM fluctuations.
The angular dependence for B$_{\rm 1g}$ and E$_{\rm u}$ are expressed as
\begin{align}
\phi_{\bm{k}}^{\rm B_{1g}} &\propto \cos k_x - \cos k_y, \\
\phi_{\bm{k}}^{\rm E_u} &\propto \sin k_{\xi} \quad (\xi=x,y),
\end{align}
which may be referred to as $d$-wave and $p$-wave, respectively. 
The transition temperature determined from the condition $\lambda_{\rm SC}=1$ is shown in Fig.~\ref{fig:FS-area}(b).
The $d$-SC occurs in the weak-coupling regime $J\lesssim 0.9$, 
while the $p$-wave superconductivity has a higher transition temperature in $J\gtrsim 0.9$.
We observed a similar behavior in the wide-doping range of $n \gtrsim 0.80$.

It is noteworthy here that the competing $d$-wave and $p$-wave superconductivities have been discussed
in a context of CeRhIn$_5$ and CeCu$_2$Si$_2$~\cite{Fuseya03}.
By a phenomenological treatment of AFM QCP, properties of the $p$-wave singlet superconductivity was investigated.

As $J$ is increased further beyond $J_{\rm c}$, on the other hand, we observed the leading $s$-wave fluctuations.
Figure~\ref{fig:SC-eigen}(b) shows $\lambda_{\rm SC}$ and $\phi_{\omega_0 \bm{k}}$ for $J=1.8$ and $n=0.92$.
The eigenfunction of the $s$-wave fluctuations has the momentum dependence expressed by
\begin{align}
\phi_{\bm{k}}^{\rm A_{1g}} &\propto \cos k_x + \cos k_y,
\end{align}
which may be referred to as an extended-$s$ wave.
The enhanced $s$-wave pairing was observed in the low-doped and strong-coupling regime, $n \gtrsim 0.88$ and $J\gtrsim J_{\rm c}$.
However, we did not find a transition to the $s$-wave superconductivity, i.e., a parameter set which gives $\lambda_{\rm SC}=1$, or could not reach low enough temperatures because of a bad convergence of the bath function $\Delta_{\omega}$.

In order to identify the driving force of the $s$-wave fluctuations, we performed DMFT calculations without dual fermions as well. In this case, no enhancement of the fluctuations was observed within temperatures we have reached~\footnote{There is a report that an $s$-wave superconductivity occurs in the Kondo lattice model within DMFT~\cite{Bodensiek13}. It employs the infinite-dimensional Bethe lattice, and cannot be strictly compared with our results.}.
Therefore, we conclude that spatial correlations are relevant for enhancement of the $s$-wave fluctuations. 
An $s$-wave superconductivity was recently reported in the periodic Anderson model~\cite{Masuda15}, 
and its connection with our results is left for future investigations.

{\em Quasiparticle band.}
We discuss quasiparticle properties to reveal the origin of the competing superconducting fluctuations.
Two contrasting spectra
$A(\bm{k}, \omega)=-\pi^{-1} {\rm Im}G_{\omega\bm{k}}$
are shown in Fig.~\ref{fig:FS-area}(a). Here, we used the Pad\'e approximation for analytic continuations.
The non-interacting dispersion $\epsilon_{\bm{k}}$ can be traced in $J=0.8$, while the spectrum exhibits a distinct hybridization gap in $J=2.6$ as a consequence of the Kondo screening.
The formation of quasiparticles can be quantified in terms of the topology of the Fermi surface (Fermi line in two dimensions)~\cite{Otsuki09a}.
To this end, we evaluated the area $n_{\rm FS}$ surrounded by the Fermi line:
\begin{align}
n_{\rm FS} = \frac{2}{N} \sum_{\bm{k}} \Theta(\mu - \epsilon_{\bm{k}} - {\rm Re} \Sigma_{\omega=0, \bm{k}}),
\label{eq:n_FS}
\end{align}
where $\Theta$ is the step function.
A crossover between the non-interacting band ($n_{\rm FS}=n$) and the heavy-fermion band ($n_{\rm FS}=n+1$) is thus visualized in the $J$-$T$ plane in Fig.~\ref{fig:FS-area}(b).
It is worth notice that the crossover region coincides with the coupling region where the change of the pairing symmetry is observed.

In what follows, we discuss the origin of the competing superconducting fluctuations in connection with the crossover of the Fermi-surface structure.
As we have pointed out, both $d$-wave and $p$-wave pairings are induced by AFM fluctuations.
The $d$-wave is in particular favorable in the weak-coupling regime because the intensities of $\phi_{\omega_0 \bm{k}}$ lie around the van Hove points, i.e. $\bm{k}_{1}=(\pi,0)$.
The advantage of the $d$-SC, however, diminishes as the hybridization band is formed.
The $p$-wave superconductivity instead emerges in the crossover region between small and large Fermi surfaces as shown in Fig.~\ref{fig:FS-area}(b).
In this region, low-energy excitations exist around $\bm{k}_{2}=(\pi/2, \pi/2)$ rather than around the van Hove points $\bm{k}_{1}$.
Therefore, the scattering between $\bm{k}_{2}$ and $-\bm{k}_{2}$ gives a major contribution to pairing interactions to favor the $p$-wave symmetry.
Furthermore, dynamical nature of quasiparticle interactions is relevant for a realization of odd-frequency pairings~\cite{Balatsky92,Fuseya03}.
In the present formalism, the vertex $\gamma$ plays its role, and it indeed has a strong frequency dependence in the heavy-fermion regime.

In summary,
the $d$-SC is not particularly favored in the Kondo lattice even though the nesting at $\bm{q}=(\pi,\pi)$ gives rise to critical AFM fluctuations.
The key feature which yields differences with the Hubbard model is the Fermi-surface crossover: The nesting of the non-interacting band governs the magnetic instability, while the hybridization band is formed over the evolution of heavy quasiparticles.
The ``incomplete'' quasiparticles in the crossover regime undergo the critical AFM fluctuations, 
resulting in emergent quantum states such as the $p$-wave spin-singlet superconductivity.

The author thanks H. Kusunose for encouraging discussions.
This work was supported by JSPS KAKENHI Grant No. 26800172.
A part of the computations was performed in the ISSP Supercomputer Center, the University of Tokyo.

\bibliography{JO,footnote}

\end{document}